\documentclass[twocolumn,aps,superscriptaddress,amsmath,showpacs,tightenlines]{revtex4}
\usepackage{epsfig,graphicx,times}
\usepackage{amstext}
\usepackage{amsmath}            %serve per le subequazioni
\usepackage{amssymb}            %serve per il simbolo "marchio registrato", \circledR
\usepackage{graphicx}           %serve per le figure eps, ps etcyy
\usepackage{latexsym}
\usepackage{bm}

\begin{document}
\title{Preparation of macroscopic quantum superposition states of a cavity field
via coupling to a superconducting charge qubit}

\author{ Yu-xi Liu}
\affiliation{Frontier Research System,  The Institute of Physical
and Chemical Research (RIKEN), Wako-shi 351-0198}
\author{L.F. Wei}
\affiliation{Frontier Research System,  The Institute of Physical
and Chemical Research (RIKEN), Wako-shi 351-0198}
\affiliation{Institute of Quantum Optics and Quantum Information,
Department of Physics, Shanghai Jiaotong University, Shanghai
200030, P.R. China }
\author{Franco Nori}
\affiliation{Frontier Research System,  The Institute of Physical
and Chemical Research (RIKEN), Wako-shi 351-0198}
\affiliation{Center for Theoretical Physics, Physics Department,
Center for the Study of Complex Systems, The University of
Michigan, Ann Arbor, Michigan 48109-1120}

\date{\today}

\begin{abstract}
We propose how to generate Schr\"odinger cat states using a
microwave cavity containing a SQUID-based charge qubit. Based on
the measurement of charge states, we show that the superpositions
of two macroscopically distinguishable coherent states  of a
single-mode cavity field can be generated by a controllable
interaction between a cavity field and a charge qubit. After such
superpositions of the cavity field are created, the interaction
can be switched off by the classical magnetic field through the
SQUID, and there is no information transfer between the cavity
field and the charge qubit. We also discuss the generation of
superpositions of two squeezed coherent states.

\end{abstract}

\maketitle \pagenumbering{arabic}
\section{Introduction}

The principle of linear superposition is central to quantum
mechanics. However, it is difficult to create and observe
superposed states because the fragile coherence of these states
can be easily spoiled by the environment. Typical examples are the
Schr\"odinger cat states (SCSs)~\cite{mb}. Many theoretical
schemes~\cite{theor} have been proposed to generate SCSs and
superpositions of macroscopic states (SMSs) in optical systems.
Also, much experimental progress has been made to demonstrate SMSs
and SCSs: in superconducting systems (e.g. Ref.~\cite{super}),
laser trapped ions~\cite{trapped}, optical systems constructed by
Rydberg atoms, and superconducting cavity in the microwave
regime~\cite{mb,opt}. The SMSs, which are formed by two optical
coherent states, e.g. in Ref.~\cite{co}, have been investigated
for applications in quantum information
processing~\cite{co,application,liuy}. These states can be used as
a robust qubit encoding  for a single bosonic mode subject to
amplitude damping. They can also be used to study both the
measurement process and decoherence by coupling the system to the
external environment~\cite{co,application,liuy}. Thus,  generating
and measuring SMSs and SCSs are not only important to understand
fundamental physics, but also to explore potential applications.

Superconducting quantum devices
~\cite{moo,Y.Nakamura,vion,you,wei} allow to perform quantum state
engineering including the demonstration of  SCSs and SMSs.
Theoretical schemes to generate superpositions that are different
from the above experiments~\cite{super} have also been
proposed~\cite{cc,fm,armour} in superconducting quantum devices.
For example, the scheme in Ref.~\cite{cc} generates a
superpositions of Bloch states for the current of a Josephson
junction. Marquardt and Bruder~\cite{fm} proposed ways to create
SMSs for a harmonic oscillator approximated by a large
superconducting island capacitively coupled to a smaller
Cooper-pair box. Armour {\it et al.}~\cite{armour} proposed a
similar scheme as in Ref.~\cite{fm} but using a micromechanical
resonator as the harmonic oscillator. A review paper on
micromechanical resonators~\cite{scc} can be found in
Ref.~\cite{mc}. In Ref.~\cite{mje}, a scheme was proposed to
generate SMSs and squeezed states for a superconducting quantum
interference device (SQUID) ring modelled as an oscillator. Since
then, several proposals have been made which focus on
superconducting qubits interacting with the nonclassical
electromagnetic
field~\cite{saidi,zhu,you,sun,vourdas,gir,wallraff}.

Optical states allow a fast and convenient optical transmission of
the quantum information which is stored in charge qubits. Compared
with the harmonic system~\cite{fm,armour} formed by the large
superconducting junction and the micromechanical resonator,
optical qubits can easily fly relatively long distances between
superconducting charge qubits. Moreover, the qubit formed by SMSs
enables a more efficient error correction than that formed by the
single photon and vacuum states, and the generation and detection
of  coherent light are easy to implement.

In contrast to~\cite{fm,armour}, here we aim at generating SCSs in
the interaction system between a single-mode microwave cavity
field and a SQUID-based charge qubit, and then creating SMSs by
virtue of the measurements of the charge states. The generation of
such states has been studied theoretically~\cite{ld} and
demonstrated in optical cavity QED experiments~\cite{opt}.
However, in these cases: i) several operations are needed because
atoms must pass through three cavities, and ii) the interaction
times are tuned by the controlling velocity of the atoms flying
through the cavity. In our proposal, we need only one cavity, and
interaction times are controlled by changing the external magnetic
field.

Although our scheme is similar to that proposed in
Ref.~\cite{armour}, the interaction between the box and the
resonator in Ref.~\cite{armour} is not switchable. Due to the
fixed coupling in Ref.~\cite{armour}, the transfer of information
between the micromechanical resonator and the box still exists
even after the SCSs or SMSs are produced. In our proposal, the
interaction between the cavity field and the SQUID can be switched
off by a classical magnetic field after the SCSs or SMSs are
generated. Furthermore, three operations, with different
approximations made in every operation, are required in
Ref.~\cite{armour}. In addition, in order to minimize the
environmental effect on the prepared state, the number of
operations and instruments should be as small as possible; {\it
one} operation  is enough to generate SCSs or SMSs. Thus our
proposed scheme offers significant advantages over the pioneering
proposals in Refs.~\cite{fm} and~\cite{armour}.

\section{model}
We consider a SQUID-type qubit superconducting box with $n$ excess
Cooper-pair charges connected to a superconducting loop via two
identical Josephson junctions with capacitors $C_{\rm J}$ and
coupling energies $E_{\rm J}$. A controllable gate voltage $V_{\rm
g}$ is coupled to the box via the gate capacitor $C_{\rm g}$ with
dimensionless gate charge $n_{\rm g}=C_{\rm g}V_{\rm g}/2e$. The
qubit is assumed to work in the charge regime with $k_{\rm B}T\ll
E_{\rm J}\ll E_{\rm C}\ll \Delta$, where $k_{\rm B}$, $T$, $E_{\rm
C}$, and $\Delta$ are the Boltzmann constant, temperature, charge
and superconducting gap energies, respectively. For known charge
qubit experiments, e.g. in Ref.~\cite{Y.Nakamura}, $T\sim 30$ mK
which means $k_{\rm B}T\sim 3\mu$eV, $E_{\rm J}\sim 51.8\mu$eV,
$E_{\rm C}\sim 117\mu$eV, and $\Delta\sim 230\mu$eV, so the above
inequalities are experimentally achievable. We consider a gate
voltage range near a degeneracy point $n_{\rm g}=1/2$, where only
two charge states, called $n=0$ and $n=1$, play a leading role.
The other charge states with a much higher energy can be
neglected, which implies that the superconducting box can be
reduced to a two-level system or qubit~\cite{y}.  This
superconducting two-level system can be represented by a
spin-$\frac{1}{2}$ notation such that the charge states $n=0$ and
$n=1$ correspond to eigenstates $|\uparrow\rangle$ and
$|\downarrow\rangle$ of the spin operator $\sigma_{z}$,
respectively. If such a qubit is placed into a single-mode
superconducting cavity, the Hamiltonian can be written
as~\cite{you,liu}
\begin{eqnarray}\label{eq:1}
H=\hbar\omega a^{\dagger}a+E_{z}\sigma_{z}-E_{\rm J}\sigma_{x}
\cos\frac{\pi}{\Phi_{0}}\left(\Phi_{c} I+\eta
\,a+\eta^{*}\,a^{\dagger}\right),
\end{eqnarray}
where the first term represents the free Hamiltonian of the
single-mode cavity field with frequency $\omega$, and
$E_{z}=-2E_{\rm ch}(1-2n_{\rm g})$ with the single-electron
charging energy $E_{ch}=e^2/(C_{\rm g}+2C_{\rm J})$. Here
$\Phi_{0}$ is the flux quantum, $\Phi_{c}$ is the flux generated
by the classical magnetic field through the SQUID, and $I$ is the
identity operator. The last term in Eq.~(\ref{eq:1}) is the
nonlinear photon-qubit interaction. The parameter $\eta$ has units
of magnetic flux and its absolute value represents the strength of
the quantum flux inside the cavity. We will later on assume this
``quantum magnetic flux" $\eta$ to be small, becoming our
perturbation parameter. The parameter $\eta$ can be written as
\begin{equation}\eta=\int_{S} \mathbf{u}(\mathbf{r})\cdot
d\mathbf{s},
\end{equation}
where $\mathbf{u}(\mathbf{r})$ is the mode function of the
single-mode cavity field, with annihilation (creation) operators
$a \,(a^{\dagger})$, and $S$ is the surface defined by the contour
of the SQUID. For convenience, hereafter, we denote
$|\!\downarrow\rangle$ and $|\!\uparrow\rangle$ by $|e\rangle$ and
$|g\rangle$, respectively. The cosine in Eq.~(\ref{eq:1}) can be
decomposed into classical and quantized parts;  Eq.~(\ref{eq:1})
can then be expressed as
\begin{eqnarray}\label{eq:2}
H&=&\hbar\omega a^{\dagger}a-E_{\rm J}\sigma_{x}
\cos\left(\frac{\pi
\Phi_{c}}{\Phi_{0}}\right)\cos\frac{\pi}{\Phi_{0}}\left(\eta
\,a+\eta^{*}\,a^{\dagger}\right) \nonumber\\
&+&E_{z}\sigma_{z}+E_{\rm J}\sigma_{x} \sin\left(\frac{\pi
\Phi_{c}}{\Phi_{0}}\right)\sin\frac{\pi}{\Phi_{0}}\left(\eta
\,a+\eta^{*}\,a^{\dagger}\right).
\end{eqnarray}
The factors  $\sin [\pi(\eta\, a+H.c.)/\Phi_{0}]$ and $\cos
[\pi(\eta\, a+H.c.)/\Phi_{0}]$  can be further expanded as a power
series in $a \,(a^{\dagger})$. For the single photon transition
between the states $|e,n\rangle$ and $|g, n+1\rangle$, if the
condition
\begin{equation}\label{eq:33}
\frac{\pi|\eta|}{\Phi_{0}}\sqrt{n+1}\ll 1
\end{equation}
is satisfied,  all higher orders of $\pi|\eta|/\Phi_{0}$ can be
neglected in the expansion of Eq.~(\ref{eq:2}). To estimate the
interaction coupling between the cavity field and the qubit, we
assume that the single-mode cavity field is in a standing-wave
form

\begin{equation*}
B_{x}=-i\sqrt{\frac{\hbar\omega}{\varepsilon_{0}V
c^{2}}}(a-a^{\dagger})\cos(k z),
\end{equation*}
where $V$, $\varepsilon_{0}$, $c$, and $k$ are the volume of the
cavity,  permittivity of the vacuum, light speed and wave vector
of the cavity mode, respectively.
Because the superconducting
microwave cavity is assumed to only contain a single mode of the
magnetic field, the wave vector $k=2\pi/\lambda$ is a constant for
the given cavity.
The magnetic field is assumed to propagate along the $z$ direction
and the polarization of the magnetic field is along the normal
direction of the surface area of the SQUID.
If the area of the SQUID is, e.g., of the order of $100 \,\,(\mu
\rm m)^2$, then its linear dimension, e.g., approximately of the
order of 10\,\,$\mu$m, should be much less than the microwave
wavelength of the cavity mode.
Thus, the mode function $u(\mathbf{r})$ can be considered to be
approximately independent of the integral area and the factor
$\cos(kz)$ only depends on the position $z_{0}$ where the qubit is
located.
So the parameter $\eta$ can be expressed as
$$|\eta|=S\sqrt{\frac {\hbar\omega}{\varepsilon_{0}V c^2}}|\cos(k z_{0})|,$$
which shows that the parameter $|\eta|$ depends on the area $S$
and the position $z_{0}$ of the SQUID, the wavelength $\lambda$ of
cavity field, and the volume $V$ of the cavity.
It is obvious that a larger $S$ for the SQUID corresponds to a
larger $|\eta|$.
If the SQUID is placed in the middle of a cavity with full
wavelength, that is, $z_{0}=L/2=\lambda/2$.
Then $kz_{0}=(2\pi/\lambda)(\lambda/2)=\pi$, the interaction
between the cavity field and the qubit reaches its maximum,  and
\begin{equation}
3.28\times 10^{-9}\,\leq\,\pi |\eta|/\Phi_{0}\,\leq \,7.38\times
10^{-5}\,\ll \,1
\end{equation}
in the microwave region with $15$ cm $\geq\lambda \geq 0.1$ cm.
%
%The linear dimension of the SQUID is taken as $10\,\mu$m, and the
%mode function $u(\mathbf{r})$ is assumed to be independent of the
%integral area because the dimension of the SQUID, $10 \,\,\mu$m,
%is much less than the microwave wavelength $\lambda$ (e.g., $1$cm
%) of the cavity mode.
%
For a half- or quarter-wavelength cavity, the condition $\pi
|\eta|/\Phi_{0}\ll 1$ can also be satisfied.
Therefore, the approximation in Eq.~(\ref{eq:33}) can be safely
made in the microwave regime, and then Eq.~(\ref{eq:2}) can be
further simplified (up  to first order in $\xi=\pi \eta/\Phi_{0}$)
as
\begin{eqnarray}\label{eq:3}
&&H_{1}=\hbar\omega a^{\dagger}a+E_{z}\sigma_{z}-E_{\rm
J}\sigma_{x}
\cos\left(\frac{\pi \Phi_{c}}{\Phi_{0}}\right)\nonumber\\
&&+ E_{\rm J}\sigma_{x} \sin\left(\frac{\pi
\Phi_{c}}{\Phi_{0}}\right)\left(\xi
\,a+\xi^{*}\,a^{\dagger}\right),
\end{eqnarray}
where $\xi$ is a dimensionless complex number with its absolute
value equal to the dimensionless quantum magnetic flux, and it is
defined by
\begin{equation}
\xi=\frac{\pi}{\Phi_{0}}\int_{S}\mathbf{u}(\mathbf{r})\cdot
d\mathbf{s}=\frac{\pi}{\Phi_{0}}\eta.
\end{equation}

\section{Generation of cat states}
We assume that the qubit is initially in the ground state
$|g\rangle=(|+\rangle+|-\rangle)/2$ where $|+\rangle(|-\rangle)$
is eigenstate of the Pauli operator $\sigma_{x}$ with the
eigenvalue $1\, (-1)$. The cavity field is assumed initially in
the vacuum state $|0\rangle$. Now let us adjust the gate voltage
$V_{\rm g}$ and classical magnetic field such that $n_{\rm g}=1/2$
and $\Phi_{c}=\Phi_{0}/2$, and then let the whole system evolve a
time interval $\tau$. The state of the qubit-photon system evolves
into
\begin{eqnarray}\label{eq:4}
|\psi(\tau)\rangle\,&=&\,\exp\{-i[\omega
a^{\dagger}a+\sigma_{x}(\Omega^{*} a
+\Omega a^{\dagger})]\tau\}|0\rangle|g\rangle\nonumber\\
&=&\frac{1}{2}[A(\Omega)|0\rangle|+\rangle+A(-\Omega)|0\rangle|-\rangle]\nonumber\\
&=&\frac{1}{2}
(|\alpha\rangle+|-\alpha\rangle)|g\rangle\nonumber\\
&+&\frac{1}{2}(|\alpha\rangle-|-\alpha\rangle)|e\rangle,
\end{eqnarray}
where the complex Rabi frequency $\Omega=\xi^{*} E_{\rm J}/\hbar$,
$A(\pm\Omega)=\exp\{-i[\omega a^{\dagger}a\pm(\Omega^{*} a +\Omega
a^{\dagger})]\tau\}$,  and a global phase factor $\exp[-i(\xi^*
E_{\rm J}/\hbar\omega)^2\sin(\omega t)+i\xi^{*2}E^2_{\rm J}
t/\hbar^2\omega]$ has been neglected. $|\pm\alpha\rangle$ denotes
coherent state
\begin{equation}
|\pm\alpha\rangle\equiv
e^{-|\alpha|^{2}/2}\sum_{n=0}^{\infty}\frac{(\pm\alpha)^n}{\sqrt{n!}}|n\rangle
\end{equation}
with $$\alpha=\frac{\xi^{*} E_{\rm J}}{\hbar\omega}(e^{-i\omega
\tau}-1).$$ In the derivation of Eq.~(\ref{eq:4}), we use the
formula $\exp[\theta
(\beta_{1}a+\beta_{2}a^{\dagger}a+\beta_{3}a^{\dagger})]
=\exp[f_{1}a^{\dagger}]\exp[f_{2}a^{\dagger}a]\exp[f_{3}a]\exp[f_{4}]$
with the relations
$f_{1}=\beta_{3}(e^{(\beta_{2}\theta)}-1)/\beta_{2}$,
$f_{2}=\beta_{2}\theta$,
$f_{3}=\beta_{1}(e^{(\beta_{2}\theta)}-1)/\beta_{2}$, and
$f_{4}=\beta_{1}\beta_{3}(e^{(\beta_{2}\theta)}-\beta_{2}\theta-1)/\beta^{2}_{2}$.
After the time interval $\tau$, we impose $\Phi_{c}=0$ by
adjusting the classical magnetic field, thus the interaction
between the charge qubit and the cavity field is {\it switched
off} (e.g., the last term in Eq.~(\ref{eq:3}) vanishes).
Eq.~(\ref{eq:4}) shows that entanglement of the qubit and the
microwave cavity field can be prepared for an evolution time
$\tau\neq 2m\pi$ with the integer number $m$, then  Sch\"odinger
cat states can be created~\cite{mb}. If the condition $e^{-i\omega
\tau}\neq 1$ is satisfied in Eq.~(\ref{eq:4}), the SMSs of the
cavity field denoted by $|{\rm sms}\rangle_{\pm}$
\begin{equation}
|{\rm sms}\rangle_{\pm}=\frac{1}{\sqrt{2\pm
e^{-2|\alpha|^{2}}}}\,\,(|\alpha\rangle\pm|-\alpha\rangle)
\end{equation}
can be obtained by measuring the charge state $|e\rangle$ or
$|g\rangle$, by using, for example, a single-electron transistor
(SET).

If we initially inject a coherent light $|\alpha^{\prime}\rangle$,
 then by using the same method as in the derivation of
Eq.~(\ref{eq:4}), we can  also obtain the entanglement of  two
different optical coherent states $|\alpha_{\pm}\rangle$ and qubit
states with the evolution time $\tau_{1}$:
\begin{eqnarray}\label{eq:5}
 |\varphi(\tau_{1})\rangle&=&\frac{1}{2}
(\exp(i\varphi)|\alpha_{+}\rangle+\exp(-i\varphi)|\alpha_{-}\rangle)|g\rangle\nonumber\\
&+&\frac{1}{2}(\exp(i\varphi)|\alpha_{+}\rangle-\exp(-i\varphi)|\alpha_{-}\rangle)|e\rangle,
\end{eqnarray}
where
$$\varphi={\rm Im} \left[\frac{\xi E_{\rm J}}{\hbar \omega}\alpha^{\prime}(1-e^{i\omega t})\right]$$
and
$$\alpha_{\pm}\,=\,\alpha^{\prime} e^{(-i\omega \tau_{1})}\pm
\kappa [1-e^{(-i\omega \tau_{1})}]$$  with
$$\kappa=\frac{\xi^{*}E_{\rm J}}{\hbar\omega}.$$

 After a time interval $\tau_{1}$, we can switch off the
interactions between the charge qubit and the cavity field by
setting $\Phi_{c}=0$ and $n_{g}=1/2$. Measuring the charge states,
we can obtain another SMSs denoted by $|{\rm SMS}\rangle$
$$|{\rm SMS}\rangle\,=\,N^{-1}_{\pm}(e^{i\varphi}|\alpha_{+}\rangle\pm e^{-i\varphi}|\alpha_{-}\rangle)$$  with
normalized constant
$$N_{\pm}=\sqrt{2\pm (e^{-i2\varphi}\langle
\alpha_{+}|\alpha_{-}\rangle+e^{i2\varphi}\langle
\alpha_{-}|\alpha_{+}\rangle)}, $$ where $\langle
\alpha_{\mp}|\alpha_{\pm}\rangle$ can be easily
obtained~\cite{scully} by the above expression of $\alpha_{\pm}$,
for example, $$\langle
\alpha_{+}|\alpha_{-}\rangle=\exp\left\{-4\kappa^{2}[1-\cos(\omega
\tau_{1})]-i2\kappa\alpha^{\prime}\sin(\omega \tau_{1})\right\},$$
here we assume that the injected coherent field has a real
amplitude $\alpha^{\prime}$. In Eq.~(\ref{eq:5}), we entangle two
different superpositions of coherent states with the ground and
excited states of the qubit. We can also entangle two different
coherent states $|\alpha_{\pm}\rangle$ with the qubit states by
applying a classical flux such that $\Phi_{c}=\Phi_{0}$. Then with
the time evolution $t=\pi/4E_{\rm J}$, we have
\begin{eqnarray}\label{eq:6}
&&|\psi(\tau_{1})\rangle= \frac{1}{2}
(e^{-i\varphi}|\alpha_{-}\rangle|g\rangle
+e^{i\varphi}|\alpha_{+}\rangle|e\rangle).
\end{eqnarray}
It should be noticed  that a global phase factor $\exp[-i(\xi^*
E_{\rm J}/\hbar\omega)^2\sin(\omega t)+i\xi^{*2}E^2_{\rm J}
t/\hbar^2\omega]$ has been neglected in Eqs.(~\ref{eq:5}) and
~(\ref{eq:6}).

From a theoretical point of view, if we can keep the expansion
terms in Eq.~(\ref{eq:2}) up to  second order in
$\xi=\pi\eta/\Phi_{0}$, we can also {\it prepare a superposition
of two squeezed coherent states, which could be used to encode an
optical qubit}~\cite{skill}. To obtain this superposition of two
squeezed coherent states, we can set $n_{\rm g}=1/2$ and
$\Phi_{c}=0$, and derive the Hamiltonian from Eq.~(\ref{eq:2}) to
get (up to second order in $\xi$)
\begin{eqnarray}
H_{2}&=&(\hbar\omega-|\xi|^{2}E_{\rm J}\sigma_{x})
a^{\dagger}a-E_{\rm J}\left(1+\frac{|\xi|^2}{2}\right)\sigma_{x}\nonumber\\
&-&E_{\rm J}\sigma_{x}\left(\frac{\xi^2}{2}
a^{2}+\frac{\xi^{*2}}{2}a^{\dagger2}\right).
\end{eqnarray}
If the system is initially in the coherent state $|\gamma\rangle$,
and if the charge qubit is in  the ground state $|g\rangle$, we
can entangle qubit states with superpositions of two different
squeezed coherent states with an evolution time $t$ as
\begin{eqnarray}\label{eq:8}
&&|\psi( t )\rangle\nonumber =\frac{1}{2}\left[e^{-i\theta t
}|\gamma, -i\frac{\xi^{*2}E_{\rm J}}{\hbar} t\rangle+e^{i\theta
t}|\gamma, i\frac{\xi^{*2}E_{\rm J}}{\hbar}t\rangle \right]|g\rangle \nonumber\\
&&+\frac{1}{2}\left[e^{-i\theta t}|\gamma, -i\frac{\xi^{*2}E_{\rm
J}}{\hbar} t\rangle-e^{i\theta t}|\gamma, i\frac{\xi^{*2}E_{\rm
J}}{\hbar} t\rangle \right]|e\rangle,
\end{eqnarray}
where
\begin{subequations}
\begin{eqnarray}
&&\theta=E_{\rm J}\left(1+\frac{|\xi|^{2}}{2}\right),\\
&&|\gamma,\,\mp i\frac{\xi^{*2}E_{\rm J}}{\hbar}t
\rangle\,=\,U_{\pm}(t)|\gamma\rangle,
\end{eqnarray}
 and
\begin{eqnarray}
U_{\pm}(t)&=&\exp\left\{-it\left(\omega\mp\frac{|\xi|^2E_{\rm J}}{\hbar}\right)a^{\dagger}a\right\}\nonumber\\
&\times&\exp\left\{\mp i\frac{E_{\rm
J}}{\hbar}\left(\frac{\xi^2}{2} a^{2}+\frac{\xi^{*2}}{2}a^{\dagger
2}\right)t\right\}.
\end{eqnarray}
\end{subequations}
Here, $|\gamma,\,\mp i \xi^{*2}E_{\rm J} t/\hbar\rangle$ denote
squeezed coherent states, and the degree of
squeezing~\cite{cave,yariv} is determined by the time-dependent
parameter $|\xi|^{2}E_{\rm J}t/\hbar$. A superposition of two
squeezed coherent states can be obtained by the measurement on the
charge qubit. However, we should note that if we keep to first
order in $|\xi|=\pi|\eta|/\Phi_{0}$  the expansions of
Eq.~(\ref{eq:2}), the interaction between the cavity field and the
charge qubit is switchable (e.g., the last term in
Eq.~(\ref{eq:3}) vanishes for $\Phi_{c}=0$). But if we keep terms
up to second order in $|\xi|$ for the expansions of
Eq.~(\ref{eq:2}), then the qubit-field coupling is not switchable.

\section{Discussions}
Our analytical expressions show how to prepare the Schr\"odinger
states for the system of the microwave cavity field and the
SQUID-based charge qubit, we further  show that the superpositions
of two macroscopically distinguishable states can also be created
by measuring the charge states. However, similarly to the optical
cavity QED~\cite{ld}, prepared superpositions of states are
limited by the following physical quantities: the Rabi frequency
$|\Omega|=|\xi| E_{\rm J}$ (which determines the quantum operation
time $t_{q}$ of two charge qubit states through the cavity field),
the lifetime $t_{\rm d}$ of the cavity field, the lifetime $T_{1}$
and dephasing time $T_{2}$ of the charge qubit, as well as the
measurement time $\tau_{\rm m}$ on the charge qubit.

\begin{figure}
\includegraphics[width=40mm]{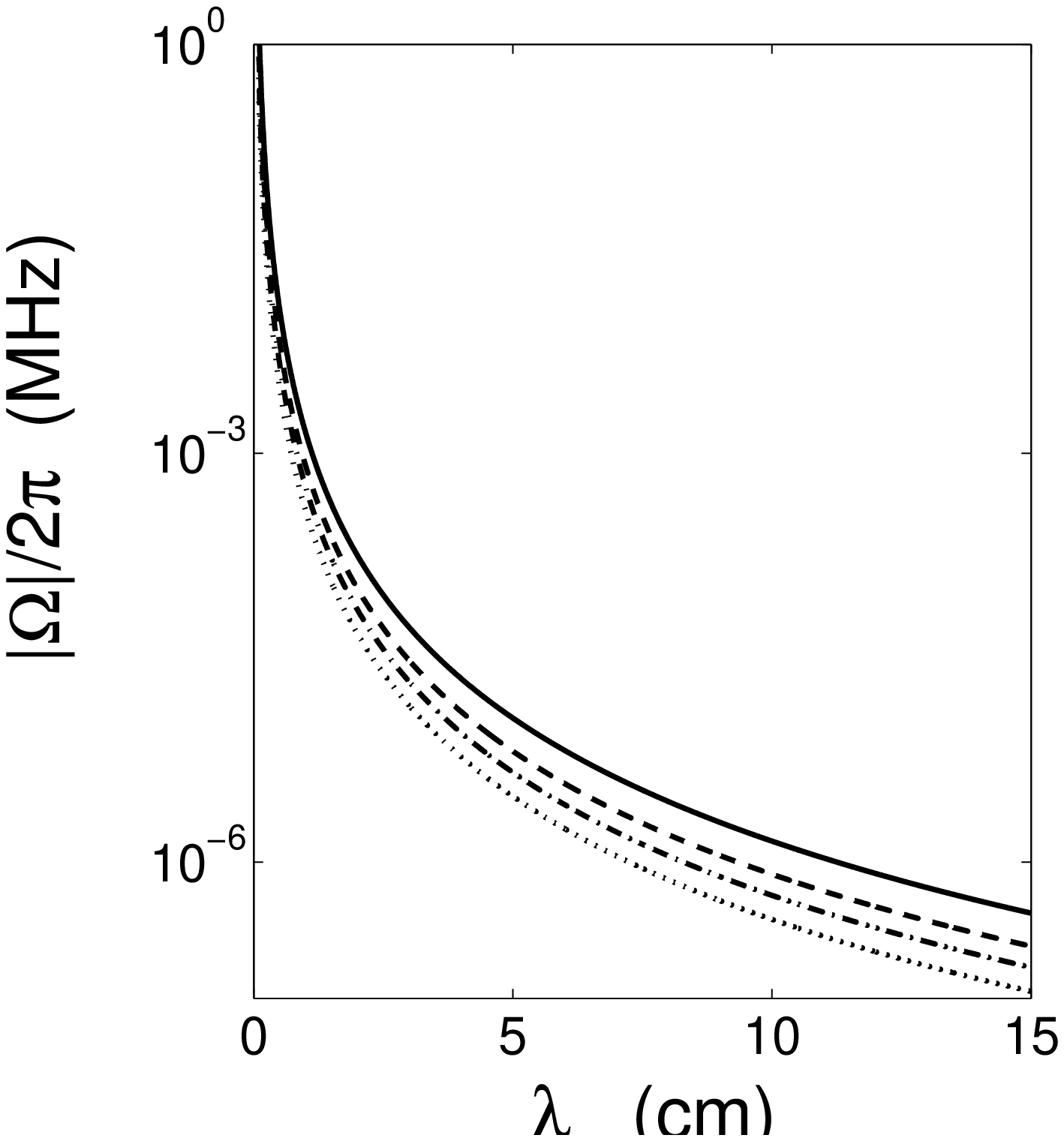}
\includegraphics[width=40mm]{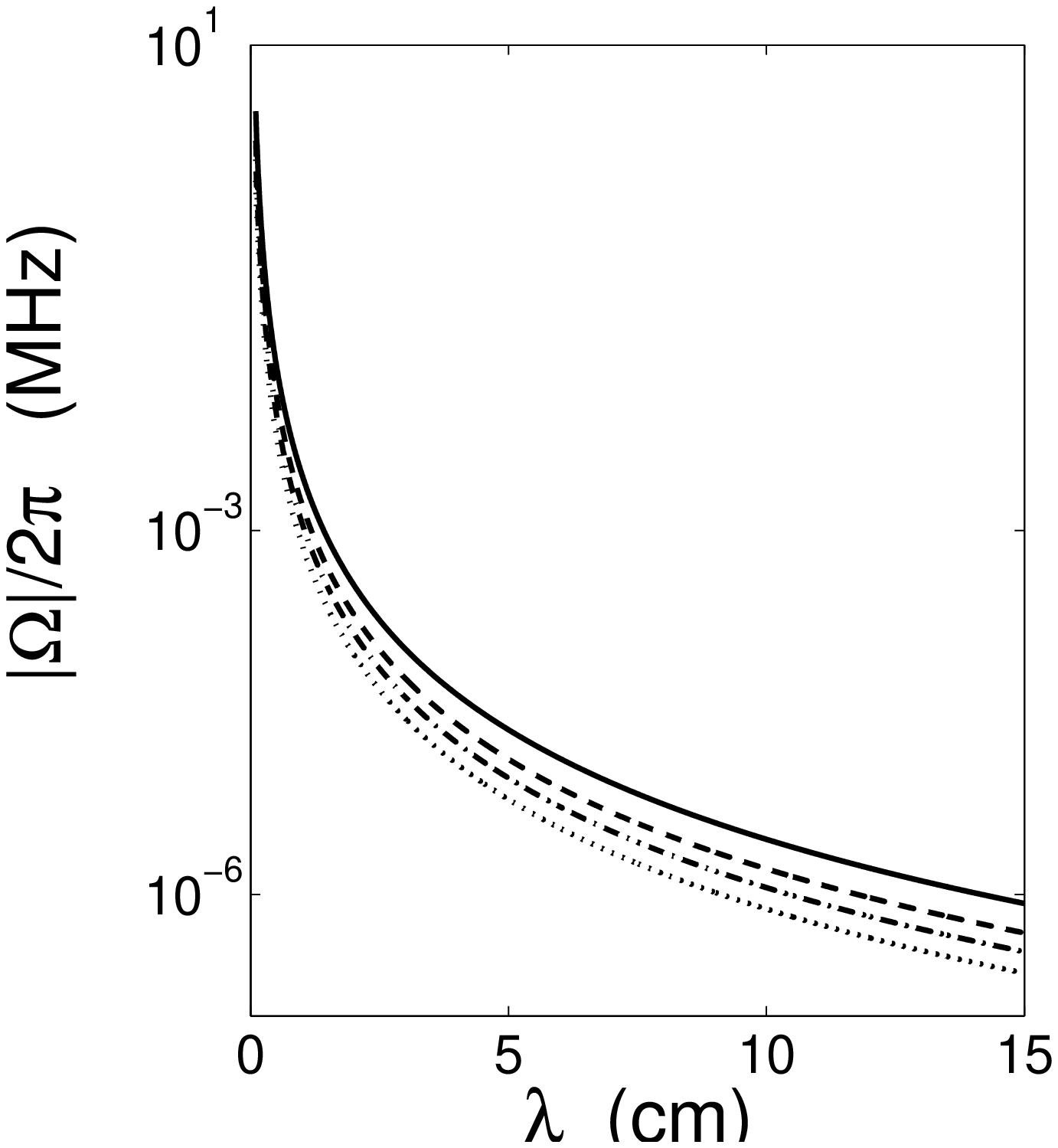}
\caption[]{Rabi frequency $|\Omega|$  versus the microwave
wavelength $\lambda$ for a full-wavelength cavity (a) and a
quarter-wavelength cavity (b) with ratios $E_{\rm ch}/E_{\rm J}=4$
({\rm top solid\, line}), 7 ({\rm dashed \,line}), 10 ({\rm
dashed-dotted \,line}),  15 ({\rm bottom dotted line}),
respectively.}\label{fig1}
\end{figure}

We now estimate the Rabi frequency $|\Omega|$ in the microwave
regime for a standing-wave field in the cavity. A SQUID with an
area of about $100 \,\,(\mu {\rm m})^2$ is assumed to be placed in
the middle of the cavity. In  the microwave regime with different
ratios of $E_{\rm ch}/E_{\rm J}$, we provide a numerical estimate
of $|\Omega|/2\pi$ for $\omega=4E_{\rm ch}/\hbar$ in a
full-wavelength cavity, shown in Fig.~\ref{fig1}(a), and  a
quarter-wavelength cavity, shown in Fig.~\ref{fig1}(b). The
results reveal that a shorter wavelength of cavity field
corresponds to a larger Rabi frequency $|\Omega|$. For example, in
the full-wavelength cavity and the case of the ratio $E_{\rm
ch}/E_{\rm J}=4$, $|\Omega|/2\pi$ with microwave length $0.1$ cm
is of the order of $10^{6}$ Hz, and yet it is about $10$ Hz for a
microwave wavelength of $5$ cm. In both cases, the transition
times from $|0\rangle|e\rangle$ to $|1\rangle|g\rangle$ are about
$10^{-6}$ s and $0.1$ s respectively, where $|0\rangle$
($|1\rangle$) is the vacuum (single-photon) state. The experiment
for this scheme should be easier for shorter wavelengths than for
long wavelengths. Since the cavity field has higher energy for the
shorter wavelength, so it is better to choose the material with a
larger superconducting energy gap to make the Josephson junction
for the experiment in the region of the shorter microwave
wavelengths. For a fixed wavelength, the effect of the ratios
$E_{\rm ch}/E_{\rm J}$ on the coupling between the cavity field
and the charge qubit is not so large. However, decreasing the
volume $V$ of the cavity can also increase the coupling.

In order to obtain a SMS, the readout time $\tau_{m}$ of the
charge qubit should be less than the dephasing time $T_{2}$ of the
charge qubit (because the relaxation time $T_{1}$ of the charge
qubit is longer than its dephasing time $T_{2}$) and the lifetime
time $t_{d}$ of the cavity field. For example,  in
Ref.~\cite{armour} with a set of given parameters, the estimated
time, $\tau_{m}=4$ ns,  is less than $T_{2}=5$
ns~\cite{Y.Nakamura}. For a good cavity~\cite{fock}, the quality
factor $Q$ can reach very high values,  such as $Q=3\times
10^{8}$, and then the lifetimes of the microwave field would be in
the range $0.001$ s $\leq 2\pi t_{d}\leq 0.15$ s, which implies
$t_{m}\ll t_{d}$. So the readout is possible within current
technology.  It is easier to prepare a SMS in such a system even
when the coupling between the charge qubit and the cavity field is
weak because, in principle, two different coherent states could be
obtained with a very short time $t_{q}$ such that $t_{q}\ll
T_{2}$. \vspace*{3cm}
\vspace{-3cm}

\section{conclusions}
In conclusion, we have analyzed the generation of Schr\"odinger
cat states via a controllable SQUID-type charge qubit. Based on
our scheme, the SMSs can be created by using one quantum operation
together with the quantum measurements on the charge qubit. After
the SCSs or SMSs are created, the coupling between the charge
qubit and the cavity field can be switched off, in principle.
Because all interaction terms of higher order in
$\xi=\pi\eta/\Phi_{0}$ are negligible for the coupling constant
$|\xi|=\pi|\eta|/\Phi_{0}\ll 1$. This results in a switchable
qubit-field interaction. This means {\em sudden switching} of the
flux on time scales of the inverse Josephson energy($>$GHz). At
present this is difficult but could be realized in the future.

We have also proposed a scheme to generate superpositions of two
squeezed coherent states if we can keep the expansion terms in
Eq.~(\ref{eq:2}) up to  second order in $\xi=\pi\eta/\Phi_{0}$.
However, in this case the interaction between the cavity field and
the charge qubit cannot be switched off. By using the same method
employed for trapped ions~\cite{qa}, we can measure the decay rate
of the SMSs and obtain the change of the $Q$ value due to the
presence of the SQUID.

Also, the generated SMSs can be used as a source of
 optical qubits. Our suggestion is that the
first experiment for generating nonclassical states via the
interaction with the charge qubit should be the generation of
superpositions of two macroscopically distinct coherent states. It
needs only one quantum operation, and the condition for the
coupling between the cavity field and the charge qubit can be
slightly relaxed. This proposal should be experimentally
accessible in the near future.

\vspace*{-0.5cm}
\section{acknowledgments}
We acknowledge comments on the manuscript from X. Hu,  S. C.
Bernstein, and C.P. Sun. This work was supported in part by the
National Security Agency (NSA), Advanced Research and Development
Activity (ARDA) under Air Force Office of Research (AFOSR)
contract number F49620-02-1-0334, and by the National Science
Foundation grant No. EIA-0130383.

\end{document}